\documentclass[aip,sd,superscriptaddress,twocolumn,a4paper]{revtex4}
\usepackage{eurosym}
\usepackage{amsmath}
\usepackage{amssymb}
\usepackage{amsbsy}
\usepackage{makeidx}
\usepackage{epsf}
\usepackage{graphicx}
\usepackage{amsfonts}
\usepackage{subfigure}
\usepackage{color}
\usepackage{vmargin}
\setmarginsrb{1.5cm}{1.5cm}{1.5cm}{2cm}{0mm}{5mm}{0mm}{10mm}
\DeclareMathAlphabet\mathbfcal{OMS}{cmsy}{b}{n}
\usepackage[all]{nowidow}
\usepackage{fancyhdr}
\fancypagestyle{title}{%
  \setlength{\headheight}{22pt}%
  \fancyhf{}
  \fancyfoot[L]{\footnotesize 1054-1500/2015/25(2)/023112/11/\$30.00}
  \fancyfoot[C]{\footnotesize {\bf 25}, 023112-\thepage}
  \fancyfoot[R]{\footnotesize \copyright \ 2015 AIP Publishing LLC}
  \fancyhead[L]{}
  \fancyhead[C]{\small CHAOS {\bf 25}, 023112 (2015)}
  \fancyhead[R]{}
}%

\begin{document}
\title{Interactions of bright and dark solitons with localized $\mathcal{PT}$%
-symmetric potentials}
\author{N.\ Karjanto}
\affiliation{Department of Mathematics, School of Science and Technology, Nazarbayev
University, Astana 010000, Kazakhstan}
\affiliation{Department of Mathematics, University College, Sungkyunkwan University, Natural Science Campus,
2066 Seobu-ro, Jangan-gu, Suwon 16419, Gyeonggi-do, Republic of Korea}

\author{W.\ Hanif}
\affiliation{School of Mathematical Sciences, University of Nottingham, University Park,
Nottingham, NG7 2RD, United Kingdom}

\author{B. A.\ Malomed}
\affiliation{Department of Physical Electronics, School of Electrical Engineering,
Faculty of Engineering, Tel Aviv University, Tel Aviv 69978, Israel}

\author{H.\ Susanto}
\email{hsusanto@essex.ac.uk}
\affiliation{Department of Mathematical Sciences, University of Essex, Wivenhoe Park,
Colchester, CO4 3SQ, United Kingdom}

\pacs{05.45.Yv; 42.65.Tg}

\begin{abstract}
\setlength{\parindent}{0pt}
{\footnotesize (Received 20 October 2014; accepted 26 January 2015; published online 23 February 2015)}\\

We study collisions of moving nonlinear-Schr\"{o}dinger solitons with a $\mathcal{PT}$-symmetric dipole embedded into the one-dimensional
self-focusing or defocusing medium. Accurate analytical results are produced for bright solitons, and, in a more qualitative form, for dark ones. 
In the former case, an essential aspect of the approximation is that it must take into regard the intrinsic chirp of the soliton, 
thus going beyond the framework of the simplest quasi-particle description of the soliton's dynamics. 
Critical velocities separating reflection and transmission of the incident bright solitons are found by
means of numerical simulations, and in the approximate semi-analytical form.
An exact solution for the dark soliton pinned by the complex $\mathcal{PT}$%
-symmetric dipole is produced too. \\
{\footnotesize \copyright \ \textsl{2015 AIP Publishing LLC.} [\url{http://dx.doi.org/10/1063/1.4907556}]}
\end{abstract}
\maketitle
\thispagestyle{title}
{\small \setlength{\parindent}{0pt}
\textbf{Collisions of moving wave packets, or solitons, bright and dark ones,
with local obstacles are problems of fundamental significance in models of
diverse linear and nonlinear wave-propagation systems. Many of such models
are based on linear and nonlinear Schr\"{o}dinger equations, which include
additional terms accounting for the local defect (obstacles).
In this context, a noteworthy fact is that collisions of nonlinear-Schr\"{o}%
dinger (NLS) solitons with local defects may lead to resonant transmission
and reflection, if the soliton's amplitude and velocity match certain
conditions. In particular, resonant excitation of a trapped mode, pinned to
the attractive defect, by the incident soliton is possible. In this work, we
make use both the linear and nonlinear Schr\"{o}dinger equations, which model
optical media and other physical systems, to consider the interaction of
incident plane waves and Gaussian wave packets in the linear model, and of
bright and dark solitons in the nonlinear one, with defects which
include both a strongly localized (delta-functional) attractive or
repulsive spatially symmetric\ (even) real potential, and an antisymmetric
(odd) imaginary part, which represents a balanced combination of
localized gain and loss. The imaginary potential of the latter type
accounts for a local term subject to the parity-time ($\mathbfcal{PT}$) symmetry,
the entire local complex defect representing a $\mathbfcal{PT}$-symmetric
\textit{dipole}},\textbf{\ embedded into the one-dimensional linear,
self-focusing, or self-defocusing medium. We use a combination of
analytical methods and simulations to consider the scattering problem in
these systems. In particular, there is a straightforward exact solution for
the scattering of plane waves in the linear model and, on the other hand, an
exact solution for dark solitons pinned to the local defect is found in the
NLS model. A basic characteristic of the scattering of bright solitons is a
critical velocity separating their reflection and transmission. 
The critical velocity is found in both numerical and approximate analytical forms. 
These theoretical results can be implemented experimentally, using the scattering of light beams on defects.}
\par}

\section{Introduction}

Losses are a ubiquitous feature appearing in all kinds of optical systems.
In most cases, losses are considered as a detrimental factor, which must be
compensated by a properly introduced gain or feeding beam, in internally and
externally driven systems, respectively~\cite{rosanov07}. However, losses may play a
positive role too, helping one to stabilize modes which otherwise would not
exist or would be unstable. An example is the possibility to stabilize dissipative solitons in
laser cavities which are described by complex Ginzburg-Landau
(CGL) equations. The simplest version of the CGL equation with the spatially
uniform linear gain and cubic loss gives rise to exact solutions in the form
of chirped sech pulses~\cite{pereira77}, but they are unstable, as the linear gain
destabilizes the zero background around the solitons. A possibility to
stabilize the solitons was proposed in Ref.~\onlinecite{malomed96}, making use of
dual-core couplers, with the linear gain acting in one core, and linear loss, 
in the other. In that system, the stable pulse exists, as an \textit{attractor}, 
along with an unstable counterpart of a smaller amplitude, which
plays the role of a separatrix between attraction basins of the stable pulse
and stable zero solution. The use of similar settings for the generation of
stable plasmonic solitons~\cite{marini11}, and for the creation of stable
two-dimensional dissipative solitons and vortices in laser systems with the
feedback described by the linearly coupled stabilizing equation~\cite{paulau10},
have been proposed too.

In this connection, it is relevant to stress a crucial difference between
dissipative solitons, which are found, in particular, in the
linearly coupled systems with the separated gain and loss~\cite{paulau10,malomed07}, and
solitons in conservative media. Stable dissipative solitons exist as
isolated attractors, selected as modes which provide for the balance between
gain and loss in the system~\cite{malomed87}. 
In addition to the phase-independent gain, such stable solitons can be supported by 
parametric amplification~\cite{bara02}. On the contrary, in conservative settings, 
including various models of nonlinear optics~\cite{KA}, solitons
exist in continuous families, rather than as isolated solutions.

More recently, a special class of dissipative systems was identified, with
exactly balanced spatially separated (antisymmetrically set) dissipative and
amplifying elements. Such systems realize the concept of the $\mathcal{PT}$
(parity-time) symmetry, which was originally elaborated in the
quantum theory~\cite{bender98} for settings described by non-Hermitian
Hamiltonians, that contain spatially even and odd real and imaginary
potentials, respectively. A distinctive feature of the Hamiltonians with
complex $\mathcal{PT}$-symmetric potentials is the fact that, up to a
certain critical value of the strength of the imaginary (dissipative) part,
their spectrum remains purely real. Such $\mathcal{PT}$-symmetric
non-Hermitian Hamiltonians of linear systems can be transformed into
Hermitian ones~\cite{mostafazadeh03}.

In terms of the quantum theory, the $\mathcal{PT}$-symmetry is a theoretical
possibility. To implement it in real settings, it is natural to resort to
the fact that the linear propagation equation for optical beams in the
paraxial approximation has essentially the same form as the Schr\"{o}dinger
equation in quantum mechanics, hence the evolution of the wave function of a
quantum particle may be emulated by the transmission of an optical beam.
This fact makes it possible to simulate many quantum-mechanical phenomena,
some of which are difficult to observe in direct experiments, by means of
relatively simple settings available in classical optics~\cite{longhi11}.

The implementation of the $\mathcal{PT}$-symmetric settings in optics, which
combines spatially symmetric refractive-index landscapes and mutually
balanced spatially separated gain and loss, was proposed in Ref.~\onlinecite{ruschhaupt05} 
and demonstrated in Ref.~\onlinecite{guo09}. These works had
drawn a great deal of attention to models of optical systems featuring the $%
\mathcal{PT}$ symmetry, see review~\cite{makris11}. A majority of such models, 
which include the Kerr nonlinearity, amount to the NLS equation for the local amplitude of the
electromagnetic wave, $\psi \left( x,z\right) $, with a complex potential,
whose real and imaginary parts, $V(x)$ and $W(x)$ are, as said above,
spatially even and odd, respectively:%
\begin{equation}
i\frac{\partial \psi }{\partial z}+\frac{1}{2}\frac{\partial ^{2}\psi }{%
\partial x^{2}}+g|\psi |^{2}\psi =\left[ V(x)+iW(x)\right] \psi .
\label{NLS}
\end{equation}%
This equation is written in terms of the spatial-domain setting, with
propagation distance $z$, the second term accounting for the the paraxial
diffraction in the transverse direction, $x$. The nonlinear term in Eq.~(\ref {NLS}) 
represents the self-focusing ($g = +1$) or defocusing ($g = -1$)
nonlinearity, in the scaled form. It was also proposed to implement the same model as the
Gross-Pitaevskii equation in Bose-Einstein condensates, with the linear gain
provided by a matter-wave laser~\cite{cartarius12}.

The presence of the nonlinearity in Eq.\ (\ref{NLS}) naturally leads to $%
\mathcal{PT}$-symmetric solitons \cite{musslimani08}, a crucially important issue being 
the stability. For $\mathcal{PT}$-symmetric couplers, 
and for models with periodic complex potentials, an accurate
stability analysis of solitons solutions was reported, respectively, in
Refs.\ \onlinecite{driben11} and \onlinecite{zezyulin11}.

Another relevant problem is wave scattering on $\mathcal{PT}$-symmetric
potentials. In particular, periodic structures can act as unidirectionally
transmitting media near the $\mathcal{PT}$-symmetry-breaking point, with
reflection suppressed at one end and enhanced at the other, as predicted
theoretically in Ref.\ \onlinecite{lin11} and demonstrated experimentally in
a metamaterial \cite{feng13}. The most natural setting for the study of the
scattering of broad linear and nonlinear wave packets (including\ solitons)
is offered by localized $\mathcal{PT}$-symmetric potentials (\textit{defects})~\cite{dmit11}. 
Such defects can be induced, for instance, by nonlinear $\mathcal{PT}$-symmetric 
oligomers embedded into a linear lattice \cite{damb12}. In the latter context, stationary states in the form of plane
waves, their reflection and transmission coefficients, and the corresponding
rectification factors, illustrating the asymmetry between left and right
propagation, were analyzed. Reflection and transmission of solitons by $%
\mathcal{PT}$-symmetric scattering potentials was studied in Ref.\ %
\onlinecite{khaw13}, where it was shown that, under special conditions, one
can have a unidirectional flow of single and multiple solitons.
Unidirectional tunneling of plane waves through optical epsilon-near-zero $\mathcal{%
PT}$-symmetric bilayers was also reported in Ref.\ \onlinecite{new}.

The subject of the present work is the interaction of linear waves and solitons, both bright and dark ones,
with a strongly localized $\mathcal{PT}$-symmetric potential, which
may be represented by the $\mathcal{PT}$ \textit{dipole}:
\begin{equation}
V(x)+iW(x)=\epsilon \delta (x)+i\gamma \delta ^{\prime }(x),  \label{delta}
\end{equation}%
where $\delta$ and $\delta^{\prime}$ denote the Dirac-delta function and its derivative, 
$\epsilon $ and $\gamma $ being real constants (positive or negative; 
note that $\gamma$ is a dimensionless parameter in the framework of the present model,
hence it may be treated as representing a small perturbation, see below, under the condition of $|\gamma| \ll 1$).
Static solutions for bright solitons pinned by the $\mathcal{PT}$ dipole
with $\epsilon <0$, which corresponds to the attractive defect, while the
host medium may be either self-focusing and defocusing, were found
in an analytical form, and their stability was investigated numerically, in
Ref.\ \onlinecite{Thawatchai}.

Previously, several techniques have been developed for analyzing
interactions of bright \cite{kivs89,hase95,KA} and dark \cite%
{KA,kivs98,fran10} solitons with inhomogeneities, such as those represented
by the complex potential in Eq.\ (\ref{NLS}). In this work, we use a
perturbation method for the consideration of interactions of moving solitons
with $\mathcal{PT}$-symmetric dipole (\ref{delta}), and report results of
systematic numerical simulations of such interactions.

The paper is organized as follows. The analytical approximation for the
bright and dark solitons are developed in Section \ref{sec2}, which also
includes a solution of the scattering problem for plane waves in the linear
medium with the embedded $\mathcal{PT}$ dipole. In that section, exact
solutions are derived too for trapped dark solitons in the model with the
self-defocusing spatially uniform nonlinearity and the $\mathcal{PT}$%
-symmetric defect (\ref{delta}). Numerical results and their comparison with
the analytical predictions are reported in Section \ref{sec3}. Conclusions
are presented in Section \ref{sec4}.

\section{Analytical considerations} \label{sec2}

\subsection{The scattering problem in the linear model}

In the linearized version of Eq. (\ref{NLS}) and (\ref{delta})
\begin{equation}
i\frac{\partial \psi }{\partial z}=-\frac{1}{2}\frac{\partial ^{2}\psi }{%
\partial x^{2}}+\left[ \epsilon \delta (x)+i\gamma \delta ^{\prime }(x)%
\right] \psi ,  \label{linear}
\end{equation}%
it is natural to consider the scattering problem for plane waves,
in the form of $\psi \left( x,z\right) =e^{ikz}U(x)$, with $k<0$ and $U(x)$
satisfying the following stationary equation:
\begin{equation}
-kU=-\frac{1}{2}U^{\prime \prime }+\left[ \epsilon \delta (x)+i\gamma \delta
^{\prime }(x)\right] U.  \label{U}
\end{equation}

The general solution of the scattering problem should be looked for as%
\begin{equation}
U(x)=\left\{
\begin{array}{lll}
e^{iqx} + \left(R_{1} + i R_{2} \right) e^{-iqx}, &\qquad& \textmd{at} \; x < 0, \\
          \left(T_{1} + i T_{2} \right) e^{ iqx}, &\qquad& \textmd{at} \; x > 0,
\end{array}%
\right.  \label{RT}
\end{equation}%
where $e^{iqx}$ with $q=\sqrt{-2k}$ represents the incident wave (arriving from the left) with the
amplitude normalized to $1$, while $\left(R_{1} + i R_{2} \right) $ and $\left(T_{1} + iT_{2}\right)$, with
real $T_{1,2}$ and $R_{1,2}$, are complex reflection and transmission
coefficients, respectively.

The boundary conditions following from Eq. (\ref{U}) at $x=0$ are%
\begin{equation}
\textmd{Jump} \left( U^{\prime }\right) =2\epsilon U_{0}, \qquad 
\textmd{Jump} \left( U\right) = 2i\gamma U_{0},  \label{jump}
\end{equation}
where Jump$(\dots)$ stands for the jump at $x = 0$, and
\begin{equation}
U_{0}\equiv \frac{1}{2}\left[ U\left( x=+0\right) +U\left( x=-0\right) %
\right]  \label{U0}
\end{equation}%
is the mean value of $U$ around $x=0$. The substitution of the generic form
of the solution to the scattering problem, in the form of Eq. (\ref{RT}),
into Eq. (\ref{jump}) yields, after some algebra, the final results
\begin{equation}
\begin{aligned}
T_{1} =  \frac{q\left( \epsilon \gamma +q\right) }{\epsilon ^{2}+q^{2}}, \qquad \qquad 
T_{2} = -\frac{q\left( \epsilon -\gamma q\right) }{\epsilon ^{2}+q^{2}},  \\
R_{1} = -\frac{\epsilon \left( \epsilon +\gamma q\right) }{\epsilon ^{2}+q^{2}}, \qquad \qquad 
R_{2} = -\frac{q\left( \epsilon +\gamma q\right) }{\epsilon ^{2}+q^{2}}. \label{TR}
\end{aligned}
\end{equation}

In particular, for $\gamma =0$, these expressions go over into
the well-known solution for the real $\delta $-functional potential:%
\begin{equation}
\begin{aligned}
T_{1} =  \frac{q^{2}}{\epsilon ^{2}+q^{2}}, &\qquad T_{2} = R_{2}=-\frac{q\epsilon}{\epsilon ^{2}+q^{2}}, \\
R_{1} = -\frac{\epsilon ^{2}}{\epsilon ^{2}+q^{2}}, & \label{gamma=0}
\end{aligned}
\end{equation}
which satisfies the \textit{unitarity condition}
\begin{equation}
T_{1}^{2}+T_{2}^{2}+R_{1}^{2}+R_{2}^{2}\equiv 1.  \label{1}
\end{equation}%
On the other hand, in the particular case of $\epsilon =0$, expressions (\ref{TR}) reduce to a simple but, apparently, novel result:%
\begin{equation}
T_{1}=1,~T_{2}=-R_{2}=\gamma ,~R_{1}=0.  \label{eps=0}
\end{equation}

Note that the general expression (\ref{TR}) and the particular one (\ref%
{eps=0}) \emph{do not }obey unitarity condition (\ref{1}), as additional
power may be generated or absorbed by the term $\sim$$\gamma$.
Indeed, expression (\ref{eps=0}) yields $%
T_{1}^{2}+T_{2}^{2}+R_{1}^{2}+R_{2}^{2}=1+2\gamma ^{2}>1$. In the general
case ($\epsilon \neq 0$), solution (\ref{TR}) produces the following result for the relative
change of the total power as the result of the scattering:%
\begin{equation}
T_{1}^{2}+T_{2}^{2}+R_{1}^{2}+R_{2}^{2}-1=\frac{2\gamma q\left( \epsilon
+\gamma q\right) }{\epsilon ^{2}+q^{2}}.  \label{change}
\end{equation}%
Thus, the scattering gives rise to the loss of the total power in the following cases (note that we fix $q>0$, while both $%
\gamma $ and $\epsilon $ may have either sign):%
\begin{equation}
\begin{aligned}
\epsilon >0, &\qquad& 0 < -\gamma <\epsilon /q;  \\
\gamma   >0, &\qquad& \epsilon <-\gamma q.  \label{loss}
\end{aligned}
\end{equation}%
Otherwise, the scattering leads to the increase of the total power.

Actually, Eq.\ (\ref{TR}), derived for the plane waves, also approximately describes
the scattering of broad pulses of finite width $l$ and central carrier wavenumber $q$, 
under condition $ql \gg 1$.

The attractive $\mathcal{PT}$ dipole, with $\epsilon <0$, gives rise to a localized pinned mode in the linear system,
\begin{equation}
U = U_{0}e^{\epsilon |x|}\left[1 + i \gamma \, \mathrm{sgn}(x)\right], \label{trapped}
\end{equation}
with arbitrary amplitude $U_{0}$, and the single eigenvalue of the
propagation constant
\begin{equation}
k = k_{0} \equiv \epsilon^{2}/2.  \label{eigen}
\end{equation}
Note that the above scattering solutions exist for $k<0$, while eigenvalue (\ref{eigen}) is positive, 
hence the pinned mode does not affect the solution of the scattering problem.
\begin{figure*}[tbh]
\subfigure[\;$v_0=1$]{\includegraphics[width=7.0cm,clip=]{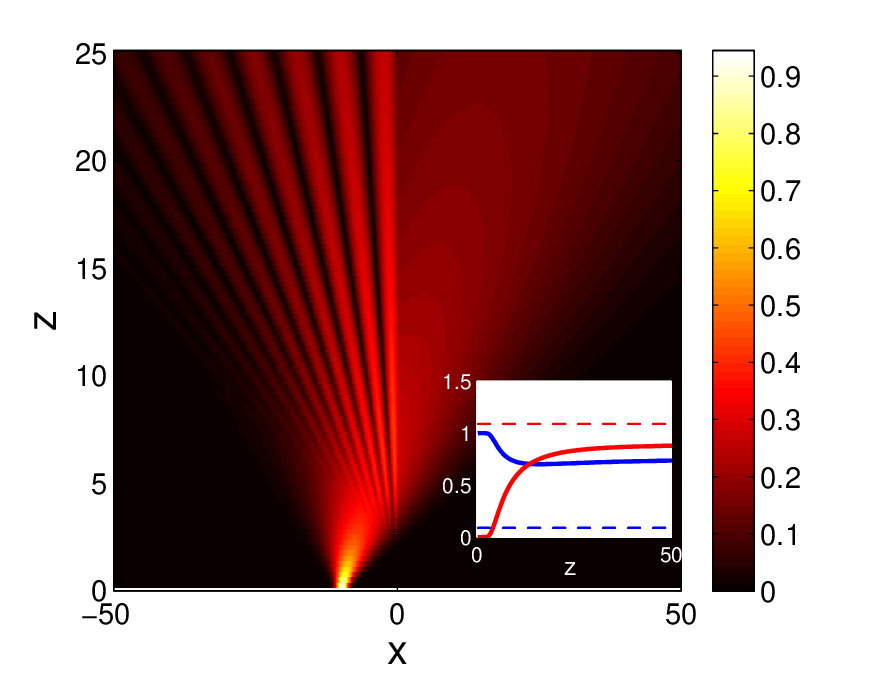}} \hspace{1cm}
\subfigure[\;$v_0=5$]{\includegraphics[width=7.0cm,clip=]{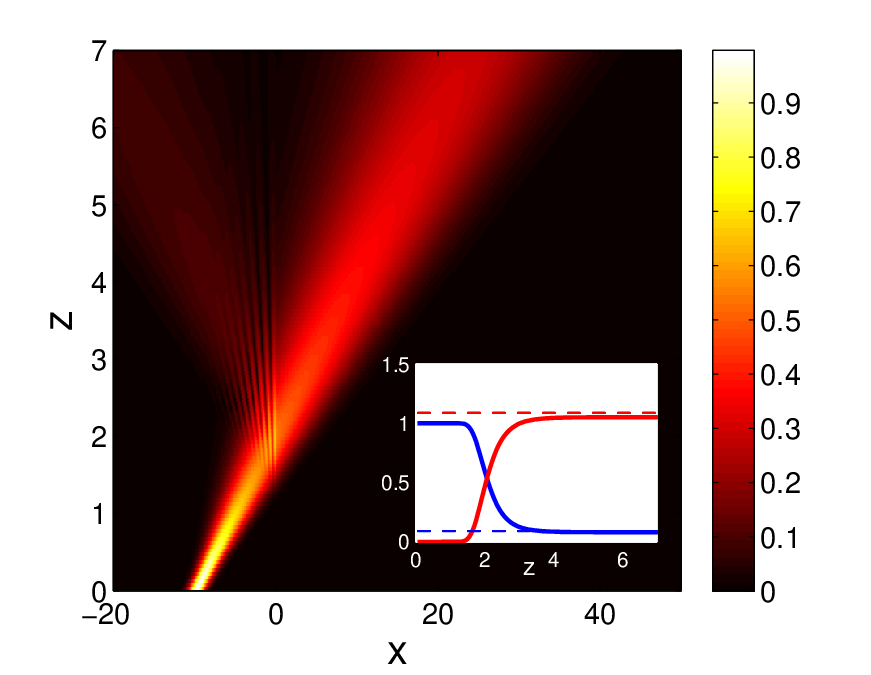}}
\caption{(Color online) The interaction of the incident Gaussian wave
packets with the localized defect in the linear model, for $\protect\epsilon =0$ and $\protect%
\gamma =0.3$. Shown is the distribution of $|\protect\psi \left( x,z\right)
| $. Solid blue and red lines in the insets depict the evolution of the
relative powers defined as $P_{R}/P_{I}$ and $P_{T}/P_{I}$, respectively,
see Eq. (\protect\ref{powers}). Their asymptotic values at $z\rightarrow
\infty $ are compared to the reflection and transmission coefficients for
the plane waves, $(R_{1}^{2}+R_{2}^{2})$ and $(T_{1}^{2}+T_{2}^{2})$, given
by Eq. (\protect\ref{TR}) (horizontal dashed lines). (a) $v_0 = 1$, (b) $v_0 = 5$.}
\label{gauss2}
\end{figure*}

\subsection{Bright solitons}

The free bright NLS soliton with amplitude $\eta $, velocity~$v$ (in fact,
it is the beam's slope in the spatial-domain setting), and coordinate $\xi$
is taken in the usual form, as the solution to Eq.\ (\ref{NLS}) with the
self-focusing sign of the nonlinearity, and $V = W = 0$
\begin{equation}
\psi \left( x,z\right) =\eta ~\mathrm{sech}\left[ \eta (x-\xi (z)\right]
\exp \left( ivx+i\phi (z)\right) ,  \label{sol}
\end{equation}%
\begin{eqnarray}
\frac{d\phi }{dz} &=&\frac{1}{2}\left( \eta ^{2}-v^{2}\right) ,  \notag \\
\frac{d\xi }{dz} &=&v.  \label{c}
\end{eqnarray}%
It is well known that the soliton may be considered as a particle with
effective mass
\begin{equation}
M=\int_{-\infty }^{+\infty }\left\vert \psi (x)\right\vert ^{2}dx\equiv 2\eta
\label{M}
\end{equation}%
and momentum
\begin{equation}
P=i\int_{-\infty }^{+\infty }\psi (x)\frac{\partial \psi ^{\ast }}{\partial x%
}dx.  \label{P}
\end{equation}%
The substitution of the unperturbed soliton's wave form$\ $(\ref{sol})
yields
\begin{equation}
P_{0} = 2 \eta v \equiv Mv. \label{P0}
\end{equation}
\begin{figure*}[tbh]
\subfigure[\;$\gamma=0.3,\,v_0=0.8$]{\includegraphics[width=7.0cm,clip=]{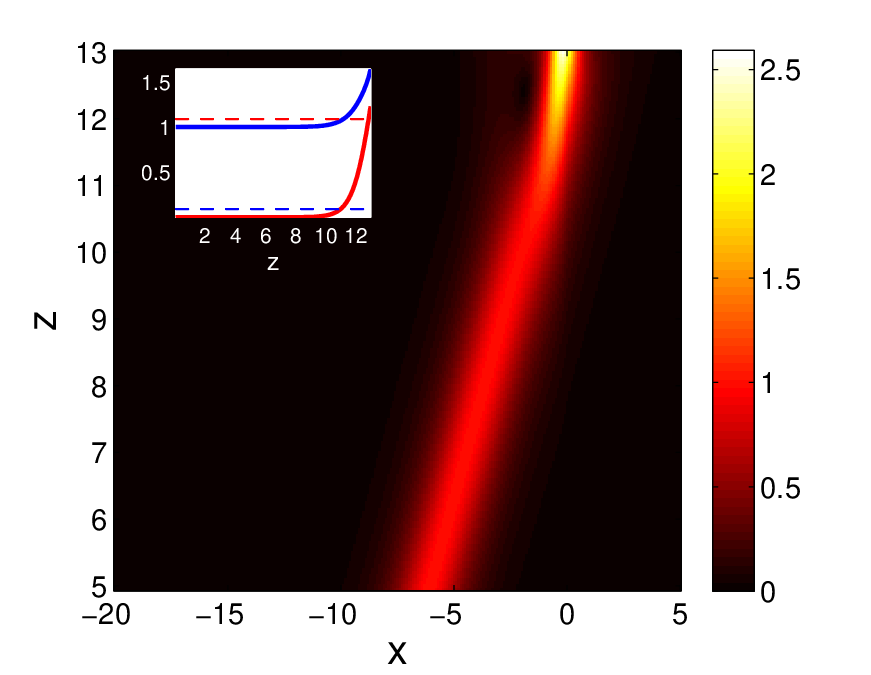}}   \hspace{1cm}
\subfigure[\;$\gamma=0.3,\,v_0=5.0$]{\includegraphics[width=7.0cm,clip=]{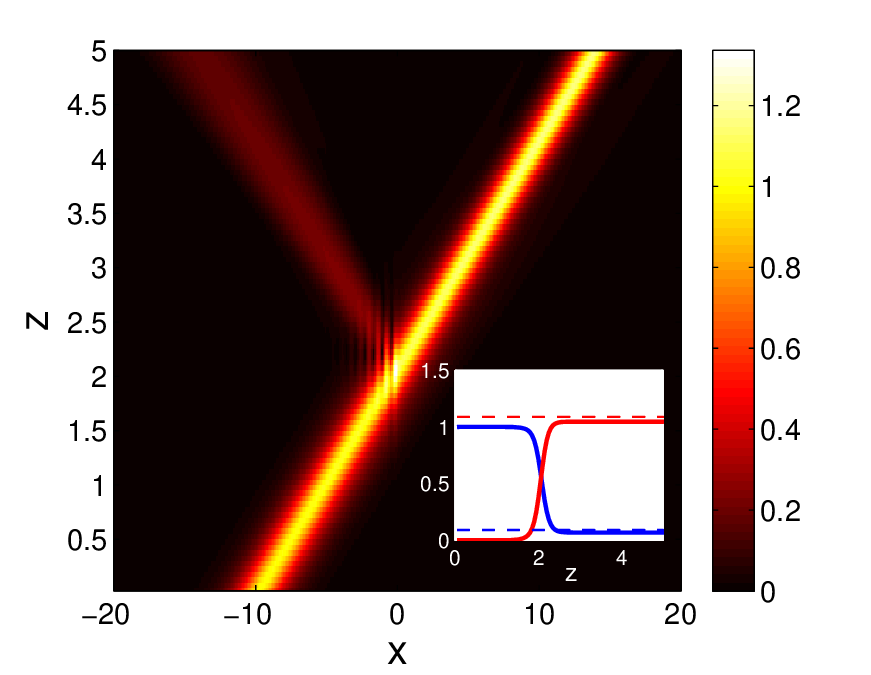}} 
\subfigure[\;$\gamma=-0.5,\,v_0=0.4$]{\includegraphics[width=7.0cm,clip=]{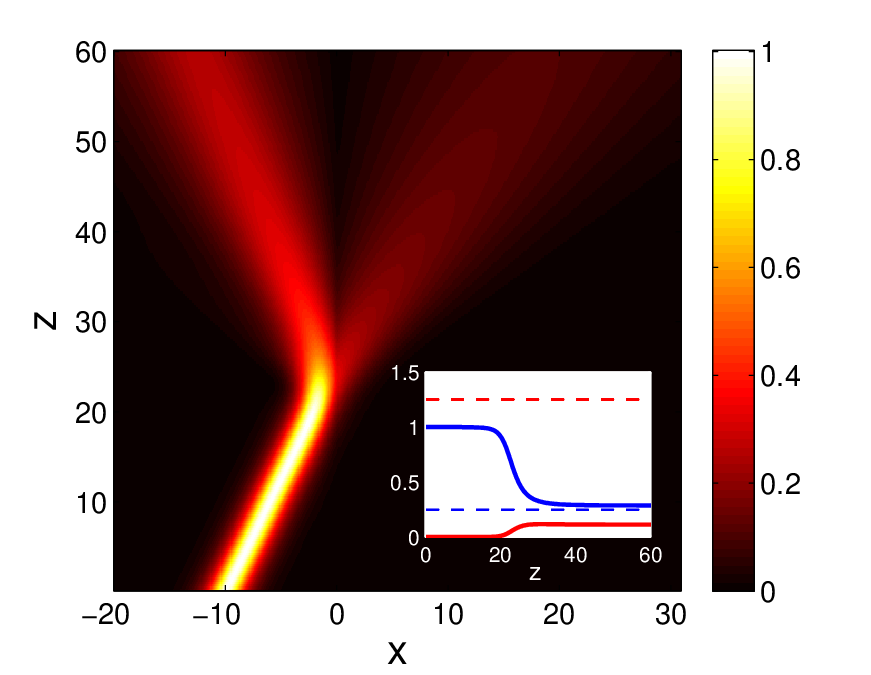}} \hspace{1cm}
\subfigure[\;$\gamma=-0.5,\,v_0=1.0$]{\includegraphics[width=7.0cm,clip=]{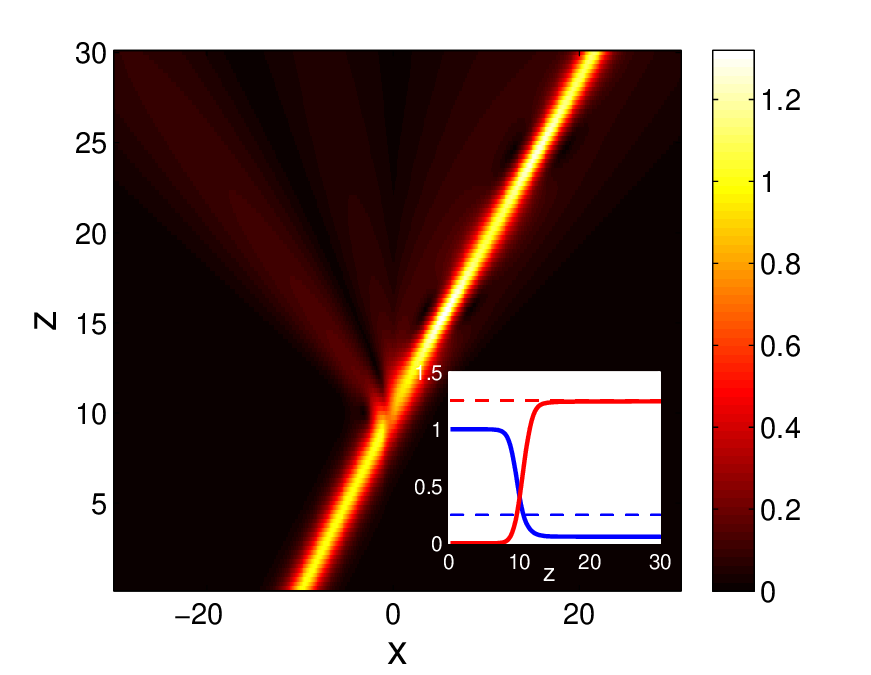}}
\caption{(Color online) Examples of the trapping and blowup (a), and
transmission (b), of the incident bright soliton interacting with the 
$\mathcal{PT}$-symmetric dipole for $\protect\gamma = 0.3$ and $\protect\epsilon =0$. In
panels (c) and (d), the incident soliton bounces back from the dipole, or
passes it, respectively, at $\protect\gamma =-0.5$ and $\protect\epsilon %
=0 $. Shown is the distribution of $|\protect\psi \left( x,z\right) |$.
Similar to Fig.\ \protect\ref{gauss2}, the insets present the evolution of
the scaled transmission and reflection powers, and compare their asymptotic
values to the respective coefficients for linear the plane waves.
(a) $\gamma = 0.3$, $v_0 = 0.8$; (b) $\gamma = 0.3$, $v_0 = 5.0$;
(c) $\gamma = -0.5$, $v_0 = 0.4$; (d) $\gamma = -0.5$, $v_0 = 1.0$.}
\label{bs1}
\end{figure*}

In the presence of Hamiltonian perturbation (\ref{delta}), with $\epsilon
\neq 0$ but $\gamma =0$, the soliton may be treated, in the adiabatic
approximation \cite{kivs89}, as a particle which keeps the constant mass ($%
d\eta /dz=0$) and moves under the action of the effective potential, $U(\xi
)=\epsilon \eta ^{2}\mathrm{sech}^{2}\left( \eta \xi \right),$ according to
Newton's equation of motion,
\begin{equation}
\frac{d}{dz}\left( 2\eta \frac{d\xi }{dz}\right) =-\frac{dU}{d\xi }%
=2\epsilon \eta ^{3}\frac{\sinh \left( \eta \xi \right) }{\cosh ^{3}\left(
\eta \xi \right) }.  \label{Newton}
\end{equation}

In the presence of the dissipative potential $\sim$$\gamma$, the mass of the
particle does not remain constant, because the total power (norm) of the
soliton evolves according to the equation%
\begin{eqnarray}
\frac{d}{dz}\int_{-\infty }^{+\infty }\left\vert \psi (x)\right\vert^{2}dx &=& 
2\int_{-\infty }^{+\infty }W(x)\left\vert \psi (x)\right\vert ^{2}dx \notag \\
&=& \left.-2\gamma \frac{\partial }{\partial x}\left( \left\vert \psi \left(x\right) \right\vert ^{2}\right) \right|_{x=0},  \label{gamma}
\end{eqnarray}%
or, after the substitution of ansatz (\ref{sol}), (which is relevant for $\gamma \ll 1$),
\begin{equation}
\frac{d\eta }{dz}=-2\gamma \eta ^{3}\frac{\sinh \left( \eta \xi \right) }{%
\cosh ^{3}\left( \eta \xi \right) }.  \label{d/dz}
\end{equation}
\begin{figure*}[tbph]
\subfigure[\;$\epsilon = 0.2$ ]{\includegraphics[width=0.35\textwidth]{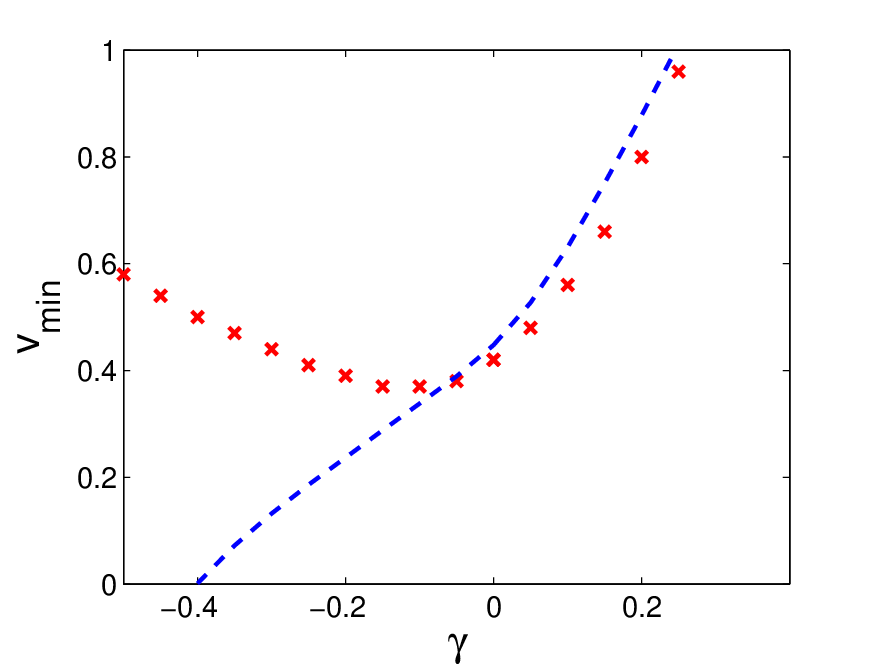}} \hspace{1cm}
\subfigure[\;$\epsilon = 0.02$]{\includegraphics[width=0.315\textwidth]{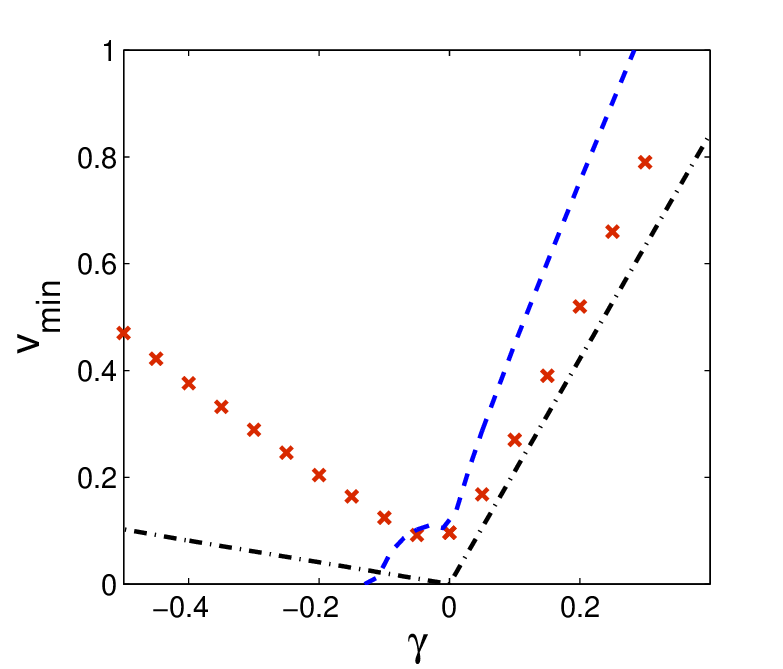}} 
\subfigure[\;$\epsilon =-0.2$ ]{\includegraphics[width=0.35\textwidth]{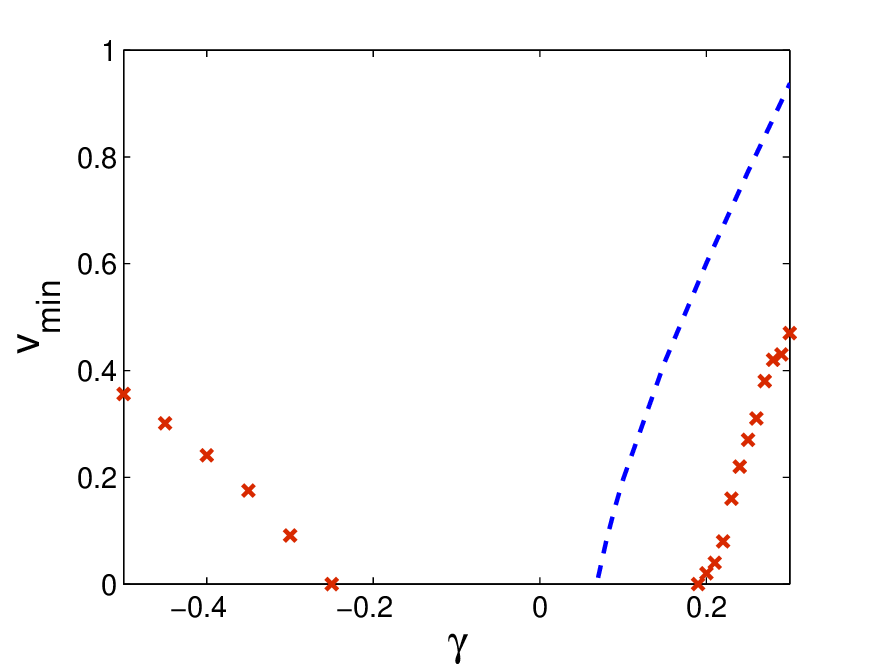}}  \hspace{1cm}
\subfigure[\;$\epsilon =-0.02$]{\includegraphics[width=0.315\textwidth]{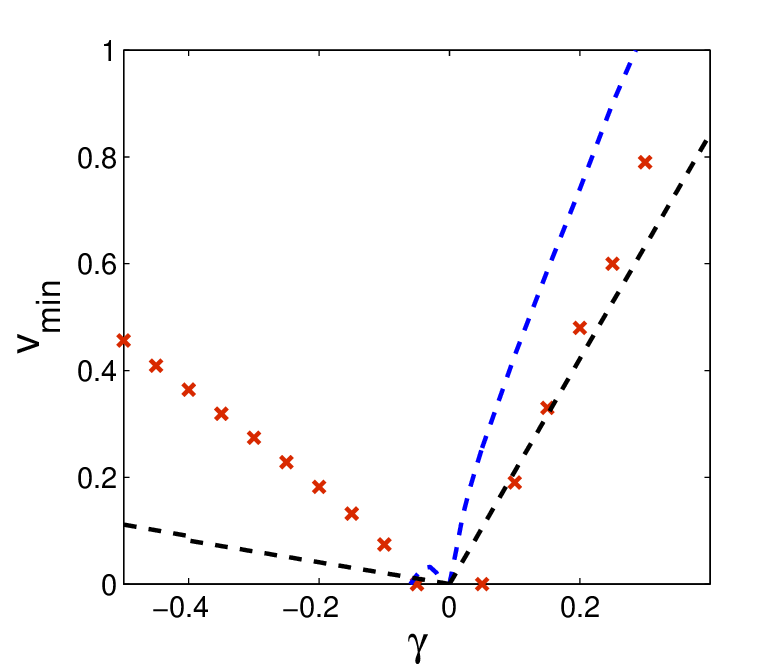}}
\caption{(Color online) The minimum velocity necessary for the transmission
of the soliton past the $\mathcal{PT}$\ dipole, which includes the local
potential, as defined by Eq. (\protect\ref{delta}). The crosses and dashed lines
represent, respectively, results of the direct simulations of Eq.\ (\protect
\ref{NLS}), and the approximation produced by a numerical solution of Eqs.\ (\protect\ref{d/dz}) and (\protect\ref{d/dz2}). For small $\protect\epsilon $
in panels (b) and (d), the approximation corresponding to Eqs.\ (\protect\ref {d/dz}) and (\protect\ref{dv/dz}) is additionally plotted by the dashed-dotted line. 
(a) $\epsilon = 0.2$, (b) $\epsilon = 0.02$, (c) $\epsilon = -0.2$, (d) $\epsilon = -0.02$.}
\label{bs3}
\end{figure*}

Under the action of the same dissipative potential, the total momentum of
the wave field, defined as in Eq.\ (\ref{P}), suffers losses according to
the equation
\begin{equation}
\left( \frac{dP}{dz}\right) _{\gamma }=\int_{-\infty }^{+\infty }W(x)\frac{%
\partial }{\partial x}\left[ \left\vert \psi (x)\right\vert ^{2}\right] dx.
\label{dP/dz}
\end{equation}
Substituting here expression (\ref{delta}) for $W(x)$ and combining it with
Newton's equation (\ref{Newton}), one arrives at the following evolution
equation:
\begin{widetext}
\begin{equation}
\frac{d}{dz}\left( \eta \frac{d\xi }{dz}\right) =\epsilon \eta ^{3}\frac{%
\sinh \left( \eta \xi \right) }{\cosh ^{3}\left( \eta \xi \right) }+\gamma
\eta ^{4}\left[ 3\mathrm{sech}^{4}\left( \eta \xi \right) -2\mathrm{sech}%
^{2}\left( \eta \xi \right) \right] ,  \label{d/dz2}
\end{equation}
\end{widetext}
where $v$ is substituted as per Eq. (\ref{c}).

Thus, the motion of the soliton interacting with the localized $\mathcal{PT}$
potential is described, in the simplest approximation, by the third-order
system of coupled ordinary differential equations (ODEs), Eqs. (\ref{d/dz}) and (\ref{d/dz2}). For the fast
incident soliton, i.e., when $d\xi /dz\left( z\rightarrow -\infty \right) = v_{0}$ is large, Eqs. (\ref{d/dz}) and (\ref{d/dz2}) can be solved
perturbatively, assuming, in the zero-order approximation,
\begin{equation}
\xi(z) = v_{0} z.  \label{c0}
\end{equation}
However, the first-order collision-induced changes of the soliton's
amplitude and momentum, $\Delta \eta$ and $\Delta \left( 2\eta v\right)$,
\emph{exactly vanish} in this limit.
Indeed, substituting approximation (\ref{c0}) into the expressions following
from Eqs. (\ref{d/dz}) and (\ref{d/dz2}),
\begin{eqnarray}
\Delta \eta &=&\int_{-\infty}^{+\infty} \frac{d\eta}{dz} dz, \nonumber \\
\Delta (2\eta c) &=& 2\int_{-\infty}^{+\infty}\frac{d}{dz}\left(\eta \frac{d\xi }{dz}\right) dz,  \label{Delta}
\end{eqnarray}
it is easy to check that both integrals are exactly equal to zero. Thus, in
the lowest-order approximation, the collision is completely elastic, which is
a manifestation of the $\mathcal{PT}$ symmetry of the model.

Numerical~results~displayed~below~[see~Fig.~\ref{bs3}(a)] demonstrate that
the full approximation, based on Eqs. (\ref{d/dz}) and (\ref{d/dz2}), is in
agreement with simulations of the underlying Eq. (\ref{NLS}) with $g = +1$
for $0 < \gamma < \epsilon $, i.e., when the local defect is composed of the
repulsive local potential and the $\mathcal{PT}$ dipole which is weaker than
the potential. When $\epsilon < 0$, i.e., the local potential is attractive,
the disagreement is anticipated [see Fig. \ref{bs3}(c) below], as the analysis
does not take into regard the formation of the trapped mode by the soliton hitting the attractive defect;
recall that, in the linear limit, the trapped mode is is given by Eq. (\ref{trapped}).

For vanishing $\epsilon $, the acceleration or deceleration of the soliton
interacting with the defect can be accounted for if the deviation of the
phase of the perturbed soliton from the adiabatic approximation,
corresponding to Eq.\ (\ref{sol}), is taken into regard. Indeed, a
well-known fact is that the perturbed soliton, whose inverse width (alias
amplitude), $\eta $, varies in the course of the evolution, $\eta =\eta (z)$%
, generates an additional \textit{chirp} term in the\ phase, hence ansatz (%
\ref{sol}) is replaced by%
\begin{eqnarray}
\psi \left(x,z\right) &=& \eta(z) \mathrm{sech}\left[\eta(z)(x - \xi(z) \right]  \notag \\
&&\times \exp \left[ ivx+ib(z)(x-\xi (z)^{2})+i\phi (z)\right], \qquad \label{chirped}
\end{eqnarray}
where, as before, the velocity is $v= d\xi/dz$, and the expression for the chirp coefficient
is produced by the variational approximation \cite{VA}
\begin{equation}
b(z)=-\left[ 2\eta (z)\right] ^{-1}\frac{d\eta }{dz}.  \label{b}
\end{equation}
Then, the substitution of the chirped ansatz (\ref{chirped}) into Eq.\ (\ref%
{dP/dz}), and the subsequent substitution of the respective correction to $dP/dz$ 
into Eq.\ (\ref{Newton}), yields, instead of (\ref{d/dz2}), a nonzero
acceleration:
\begin{eqnarray}
\frac{dv}{dz} &=& 2b\eta \int_{-\infty}^{+\infty} W(x) \frac{\left(x - \xi \right) dx}{\cosh^{2} \left[\eta(x - \xi) \right]}  \notag \\
&=& 4\gamma ^{2}\eta^{3} \frac{\tanh\left(\eta \xi \right)}{\cosh ^{4}(\eta \xi)}\left[ 2\eta \xi \tanh \left( \eta \xi \right) -1 \right], \label{dv/dz}
\end{eqnarray}
where we have inserted $W(x)=2\gamma \delta ^{\prime }(x)$, as per Eq.\ (\ref%
{delta}), expression (\ref{b}) for $b$, and Eq.\ (\ref{d/dz}) for $d\eta /dz$.

This approximation for the dynamics of bright solitons is completely
different from that derived in Ref\ \onlinecite{khaw13} for another
localized $\mathcal{PT}$-potential. Comparison of predictions based on Eqs.\
(\ref{d/dz}) and (\ref{d/dz2}) or (\ref{dv/dz}) with numerical findings is 
presented below in Section~\ref{sec3}. In particular, the \textit{post-adiabatic} 
approximation, which makes use of Eq. (\ref{dv/dz}), is accurate enough
for $\gamma >0$ and negligibly small $\epsilon $, see Figs. \ref{bs3}(b) and \ref{bs3}(d) below.

\subsection{Moving dark solitons}

Dark solitons are produced by the following modification of Eqs.\ (\ref{NLS}) and (\ref{delta}):
\begin{eqnarray}
i\frac{\partial \psi}{\partial z} &=& -\frac{1}{2}\frac{\partial^{2} \psi}{\partial x^{2}} + 
\left[\epsilon \delta(x) + i \gamma \delta^{\prime}(x) \right] \psi \nonumber \\ 
& & + \left(|\psi|^{2} - \mu \right) \psi,  \label{dark}
\end{eqnarray}
where $\mu$ is the chemical potential (i.e., squared amplitude) of the
continuous wave background maintaining the dark-soliton solution, with accordingly
defined boundary conditions at edges of the integration domain.
The self-defocusing sign of nonlinearity makes the background stable, even if it slowly varies due to the dynamics around $x = 0$.
Asymptotic theories for slowly moving dark solitons have been developed previously \cite%
{Anglin,fran02,kono04,peli08,kivs94,kivs95,kono97}. Here, we aim to present
an approximate perturbation theory for the simplest case of a moving shallow (light-gray)\ dark soliton
interacting with the $\mathcal{PT}$-symmetric dipole. Comparison of the
analysis with numerical results is not straightforward, as the simulations,
reported in Section~\ref{sec3}C, demonstrate the generation of additional
dark solitons, which is a clearly nonperturbative effect. Nevertheless, some
qualitative comparison will be possible.

We start by substituting into Eq.\ (\ref{dark}) the Madelung form,
$\psi \left(x,z\right) = \rho \left(x,z\right) \exp \left(i\phi \left(x,z\right) \right)$ 
replacing Eq.\ (\ref{dark}) by a system of real equations for the amplitude
and phase:
\begin{eqnarray}
\frac{\partial \rho }{\partial z} &=&-\frac{1}{2}\rho \frac{\partial
^{2}\phi }{\partial x^{2}}-\frac{\partial \rho }{\partial x}\frac{\partial
\phi }{\partial x}+\epsilon \delta (x)\rho ,  \label{rho} \\
\frac{\partial \phi }{\partial z} &=&\frac{1}{2}\rho ^{-1}\frac{\partial
^{2}\rho }{\partial x^{2}}-\frac{1}{2}\left( \frac{\partial \phi }{\partial x}\right) ^{2} 
-\gamma \delta ^{\prime }(x)-\left( \rho ^{2}-\mu \right). \qquad
\label{phi}
\end{eqnarray}

As in the case of Eq. (\ref{dv/dz}), we focus on the case when only the
imaginary potential is present, i.e., $\epsilon =0$ (the dynamics of dark
solitons in the presence of various real potentials was studied in detail
before in the above-mentioned works), while term $\gamma \delta^{\prime }(x)$ 
in Eq.\ (\ref{phi}) may be treated as a small perturbation.
Then, the usual approach to the description of simplest \textit{shallow} dark solitons
proceeds by setting, as suggested by \cite{tsuzuki71,bara88,shallow}
\begin{gather}
\rho = \sqrt{\mu} \left(1 + \alpha \rho_{1} \right),  \label{expansion} \\
X \equiv 2\sqrt{\alpha} \left(x + \sqrt{\mu} z \right), \qquad 
Z \equiv  \sqrt{\mu} \alpha^{3/2}z,  \label{scaling}
\end{gather}
where $\alpha$ is a formal small parameter accounting for the
shallowness of the gray soliton. The result of the analysis in the case of $\gamma = 0$ 
is the relation between the phase and amplitude perturbation, $\phi$ and $\rho_{1}$, 
$\partial \phi /\partial X = -\rho_{1}/\left( 2\sqrt{\mu }\right) $, and the Korteweg--de Vries (KdV)
equation for the evolution of the amplitude perturbation:
\begin{equation}
\frac{\partial \rho _{1}}{\partial Z}-6\rho _{1}\frac{\partial \rho _{1}}{%
\partial X}+\frac{\partial ^{3}\rho _{1}}{\partial X^{3}}=0.  \label{KdV}
\end{equation}%
At the next order, via transformations (\ref{scaling}), the perturbation
term $\gamma \delta ^{\prime }(x)$ in Eq.\ (\ref{rho}) gives rise to the
corresponding perturbation dipole term in Eq.\ (\ref{KdV}):%
\begin{equation}
\frac{\partial \rho_{1}}{\partial Z} - 6\rho_{1}\frac{\partial \rho_{1}}{\partial X} + 
\frac{\partial^{3} \rho _{1}}{\partial X^{3}} = \frac{4\gamma}{\alpha^{3/2}} \delta^{\prime} \left(X-\frac{2}{\alpha} Z \right).
\label{dipole}
\end{equation}
The term on the right-hand side of Eq.~\eqref{dipole} may be treated as a small perturbation for localized excitations 
whose amplitude $\rho_1^{(0)}$ and width $X^{(0)}$ satisfy condition
\begin{equation}
\gamma/\alpha^{3/2} \ll \rho_1^{(0)}/X^{(0)}. \label{conditiongammaalpha}
\end{equation}
Then, Eq.~\eqref{dipole} is tantamount to the perturbed KdV equation studied in Ref.~\cite{old}.
The fact that velocity of the source Eq.~\eqref{dipole} explicitly depends on $\alpha$ implies that the
present version of the perturbation theory is not a rigorously consistent one, rather offering a
qualitative analysis; in any case, it is seen below that the comparison with numerical findings for dark solitons
is possible only in a qualitative form too.
\begin{figure*}[tbph]
\subfigure[\ $\gamma = 0.50$, $\epsilon =   0$]{\includegraphics[width=0.33\textwidth]{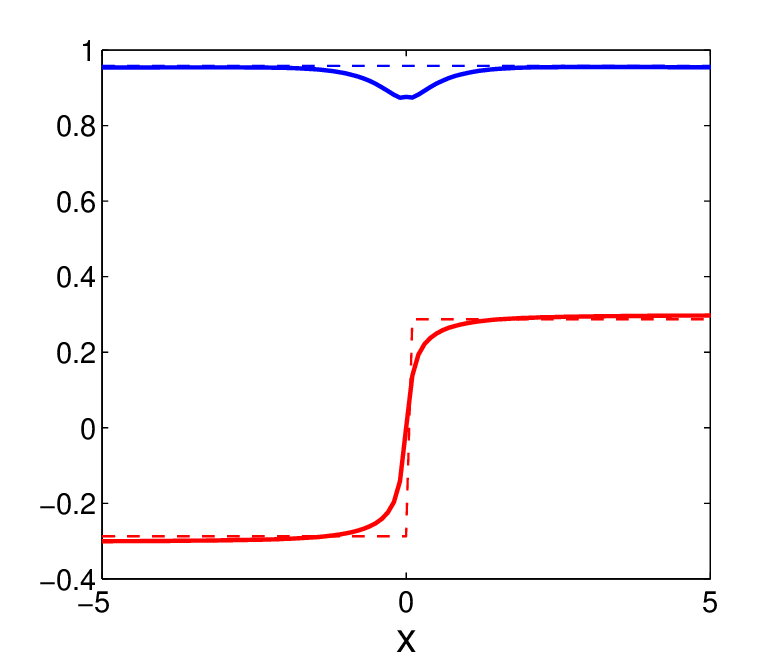}} \\
\subfigure[\ $\gamma = 0.15$, $\epsilon = 0.5$]{\includegraphics[width=0.35\textwidth]{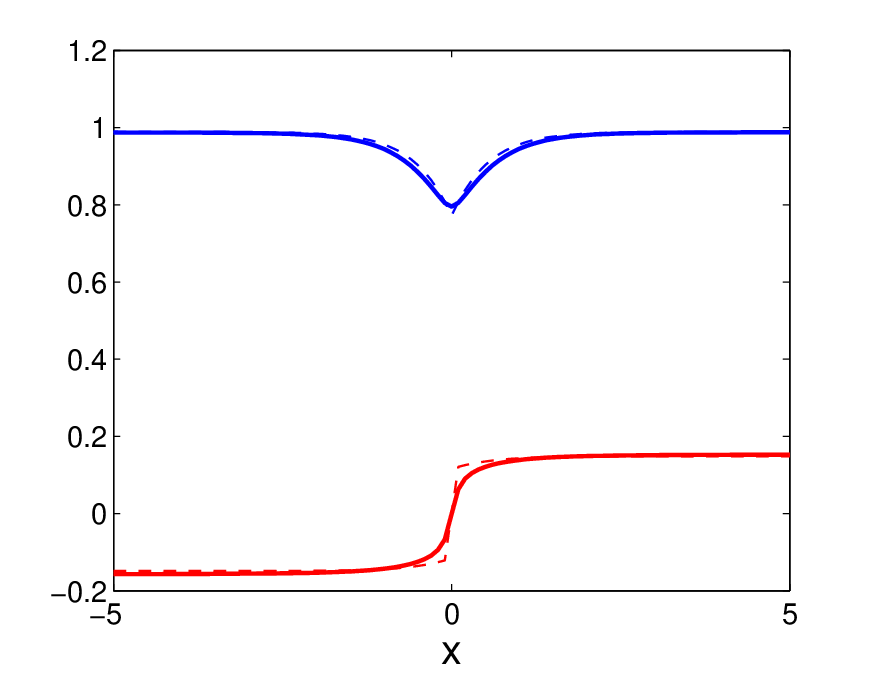}} \hspace{1cm}
\subfigure[\ $\gamma = 0.15$, $\epsilon =-0.5$]{\includegraphics[width=0.35\textwidth]{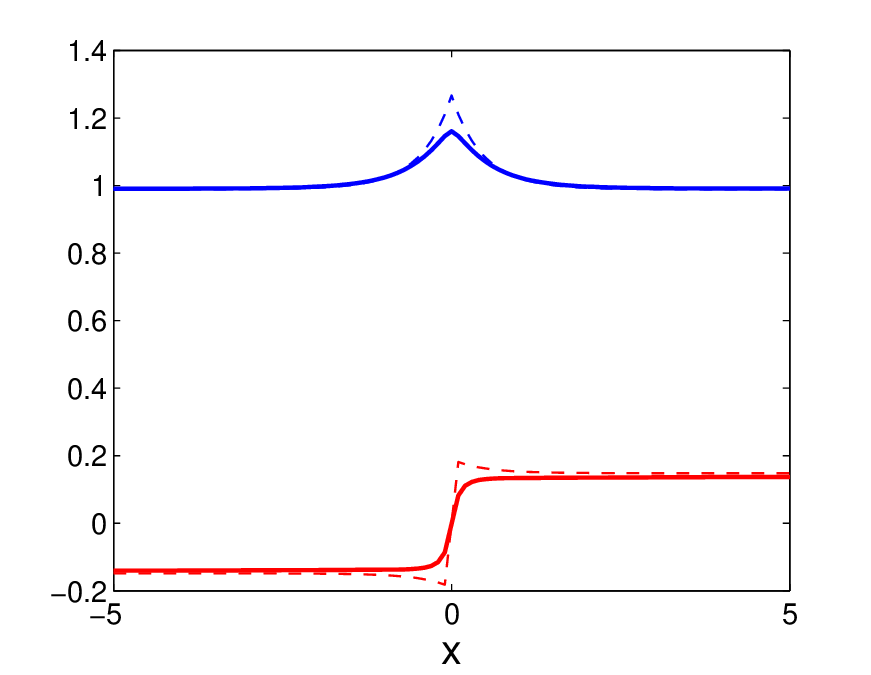}}
\caption{(Color online) (a) The numerically generated CW ground state
produced by Eq.\ (\protect\ref{dark}) with $\epsilon = 0$, $\mu = 1$, and 
the $\delta \prime$ function replaced by regularized expression~\eqref{deltas}, 
in the absence of the dark soliton. Real and imaginary parts of the solution, 
$\psi(x)$, are shown by solid lines. The dashed lines represent the analytical solution, 
which reduces to the phase jump (\protect\ref{jump}) in the case of $\protect\epsilon =0$. 
(b), (c) The same as panel (a), but for $\protect\epsilon \neq 0$ as indicated.
(a) $\gamma = 0.50$, $\epsilon = 0$; (b) $\gamma = 0.15$, $\epsilon = 0.5$;
(c) $\gamma = 0.15$, $\epsilon = -0.5$.}
\label{bg}
\end{figure*}

As shown in Ref.~\cite{old}, solutions to Eq.~(\ref{dipole}) in the form of the KdV soliton
(which represents shallow dark solitons in the present setting) interacting with the
moving dipole can be looked for as
\begin{equation}
\rho_{1} = -\frac{2 \kappa^{2}}{\cosh^{2}\left(\kappa \left(X - 2Z/\alpha \right) + \zeta(Z) \right)},  \label{solitobn}
\end{equation}
where the soliton's amplitude, $\kappa (Z)$, and position shift, $\zeta (Z)$
, evolve according to the following equations:%
\begin{eqnarray}
\frac{d\kappa}{dZ} &=& \frac{2\gamma}{\alpha^{3/2}} \frac{\kappa \sinh \zeta}{\cosh^{3} \zeta},  \label{kappa} \\
\frac{d\zeta }{dZ} &=& \kappa \left(4\kappa^{2} - \frac{2}{\alpha} \right) + \frac{2\gamma}{\alpha^{3/2}}\frac{1}{\cosh^{2}\zeta}.
\label{zeta}
\end{eqnarray}
The substitution of $\rho_1^{(0)} = \kappa^2$ and $X^{(0)} = 1/\kappa$, as per Eq.~\eqref{solitobn},
into condition~\eqref{conditiongammaalpha} casts it into the form of $\gamma/\alpha^{3/2} \ll \kappa^3$.

It was demonstrated in Ref.\ \onlinecite{old} that dynamical system (\ref{kappa}), (\ref{zeta}) 
gives rise to unbounded and trapped trajectories in the $\left(\zeta ,\kappa \right) $ plane, which, in terms of Eq.\ (\ref{dark}),
correspond, respectively, to solutions for freely moving shallow dark solitons and those trapped
by the $\mathcal{PT}$ dipole. As mentioned above, comparison of these results with numerical
simulations is possible in a qualitative form, as shown below in Section~\ref{sec3}.

\subsection{Exact solutions for pinned dark solitons}

Stationary solutions to Eq.\ (\ref{dark}) for pinned dark solitons can be
looked for as
\begin{equation}
\psi \left( x\right) =a(x)+ib(x),  \label{psi}
\end{equation}%
with $\psi (x)$ satisfying the stationary version of Eq.\ (\ref{dark}) at $%
x\neq 0$,%
\begin{equation}
-\frac{1}{2}\psi ^{\prime \prime }+\left( |\psi |^{2}-\mu \right) \psi =0,
\label{stat}
\end{equation}%
where the prime stands for $d/dx$ [no special condition like Eq.~\ref{conditiongammaalpha} is adopted here]. 
Equation (\ref{stat}) is supplemented by
the following boundary conditions at $x=0$:
\begin{eqnarray}
\textmd{Jump}(b) &=& 2\gamma a \, (x=0),  \label{D} \\
\textmd{Jump}(a^{\prime}) &=& 2\epsilon a \, (x=0),  \label{D'}
\end{eqnarray}
where Jump$\left(\dots\right)$ again stands for the jump of the respective
function at $x = 0$, cf. Eq.~\eqref{jump}. It is implied that functions $a(x)$ and $b(x)$ in
solution (\ref{psi}) are even and odd functions of $x$, respectively, hence $%
b(x=0)=0$. The corresponding solutions to Eq.\ (\ref{stat}) are found in two
different forms, depending on the sign of $\epsilon $, \textit{viz}.,\
\begin{eqnarray}
\psi (x)&=&\sqrt{\mu }\left[ \cos \theta +i~\mathrm{sgn}(x)\sin \theta \right]\nonumber\\
&&\times\tanh \left[ \sqrt{\mu }\left( |x|+\xi \right) \right] ,  \label{psipsi1}
\end{eqnarray}
for $\epsilon >0$ (the repulsive~dipole), and
\begin{eqnarray}
\psi (x) &=& \sqrt{\mu }\left[ \cos \theta +i~\mathrm{sgn}(x)\sin \theta \right]\nonumber\\
&&\times\coth \left[ \sqrt{\mu }\left( |x|+\xi \right) \right],  \label{psipsi2}
\end{eqnarray}
for $\epsilon <0$ (the attractive~one). In fact, solution (\ref{psipsi2})
describes an \textit{antidark} soliton pinned to the $\mathcal{PT}$ dipole.
The substitution of expressions (\ref{psipsi1}) and (\ref{psipsi2}) into
Eqs.\ (\ref{D}) and (\ref{D'}) yields a result which is valid for either
sign of $\epsilon$:
\begin{equation}
\begin{aligned}
&\xi = \frac{1}{2\sqrt{\mu}} \ln \left(\sqrt{\frac{4\mu}{\epsilon^{2}} + 1} + \frac{2\sqrt{\mu}}{\left\vert \epsilon \right\vert}\right), \\
&\theta = \arctan \gamma.  \label{xi}
\end{aligned}
\end{equation}
In the system with $\epsilon =0$, Eq.\ (\ref{xi}) yields $\xi =\infty$, and the corresponding solutions (\ref{psipsi1}) and (%
\ref{psipsi2}) degenerate into a constant-amplitude continuous wave (CW)
with an embedded phase jump at $x=0$,
\begin{equation}
\Delta \phi =2\arctan \gamma .  \label{shift}
\end{equation}

The solutions given by Eqs.\ (\ref{psipsi1})--(\ref{xi}) are dark-soliton
counterparts of the exact stable solutions for pinned bright solitons found
in Ref.\ \onlinecite{Thawatchai}, for $\epsilon <0$ (the attractive dipole)
and both the self-focusing and defocusing signs of the nonlinearity in Eq.\ (%
\ref{NLS}). In the limit of $\epsilon =0$, the latter solution for the
focusing nonlinearity amounts to the usual bright soliton with the same
embedded phase jump (\ref{shift}).

\section{Numerical results}

\label{sec3}

To study the soliton scattering by the $\mathcal{PT}$-symmetric dipole, we
implemented the fourth-order Runge-Kutta method for integrating Eq.\ (\ref {NLS}), 
with the Laplacian approximated by the three-point central discretization.
The simulations were carried out in spatial interval $(-L,L]$ with $L\geq 50$, 
and discrete stepsizes $\Delta x=0.1\,$and $\Delta z=0.005$ or smaller (it
was checked that further decrease of $\Delta x$ and/or $\Delta z$ did not
produce any conspicuous effect). Following Ref. \onlinecite{Thawatchai}, the
delta-function and its derivative were approximated by
\begin{equation}
\delta(x) = \frac{s}{\pi \left(x^{2} + s^{2} \right)}, \qquad 
\delta ^{\prime }(x) = -\frac{2\,s\,x}{\pi \left(x^{2} + s^{2}\right)^{2}},  \label{deltas}
\end{equation}
with $s = 0.1$. This choice secured the inner width of the regularized delta-functions to be
much smaller than the width of the incident soliton.

\subsection{Scattering of Gaussian wave packets}

First, we consider the passage of dispersive Gaussian wave packets of width 
$A^{-1/2}$ and velocity (spatial tilt) $v_{0}$ through the localized defect
in the linear system, with $g = 0$ in Eq. (\ref{NLS}). To this end, the
initial condition is taken as
\begin{equation}
\psi (x,0)=A\,e^{-A(x-x_{0})^{2}}e^{iv_{0}(x-x_{0})},  \label{gauss0}
\end{equation}
where the amplitude $A$ is fixed arbitrarily, as the model is currently
linear, and the initial position of the packet is $x_{0} = -10$.
To provide a quantitative description of the reflection and transmission, we
computed the relative powers of the wave field before the defect (at $x<0$), 
and for the field which has been transmitted past the defect (at $x>0$), $%
P_{R}/P_{I}$ and $P_{T}/P_{I}$, according to the following definitions:
\begin{equation}
\begin{aligned}
\displaystyle P_{R} = \int_{-L}^{0}|\psi (x,z)|^{2}\,dx,   \\
\displaystyle P_{T} = \int_{0}^{L} |\psi (x,z)|^{2}\,dx,   \\
\displaystyle P_{I} = \int_{-L}^{L}|\psi (x,0)|^{2}\,dx. 
\end{aligned} \label{powers}
\end{equation}%
It is natural to compare their asymptotic values at $z\rightarrow
\infty $ with the reflection and transmission coefficients for the plane
waves, $(R_{1}^{2} + R_{2}^{2})$ and $(T_{1}^{2} + T_{2}^{2})$, as given by Eq. (\ref{TR}), 
where the wavenumber $q$ is replaced by incident velocity $v_{0}$.

In Fig.~\ref{gauss2}, we display the evolution of the incident Gaussian wave
packet impinging onto the defect with $\epsilon =0$, $\gamma =0.3$, at two
different values of $v_{0}$. Shown is the top view of the absolute value of
the field, $|\psi (x,z)|$. Insets to the same figure present coefficients $%
P_{R}/P_{I}$ and $P_{T}/P_{I}$, as described above. Naturally, 
larger incoming velocity $v_{0}$ makes the values of the
coefficients at $z\rightarrow \infty $ closer to exact results for the plane
waves given by Eq.~(\ref{TR}), as the parameter accounting for the
difference of the Gaussian pulse~(\ref{gauss0}) from the plane wave is 
the ratio of the carrier wavelength to the pulse's width, $\sim$$\sqrt{A}/v_{0}$. 
The case of $\gamma < 0$ is not shown here separately, as the respective
results are quite similar to those presented in Fig.\ \ref{gauss2}.
\begin{figure*}[tbph]
\subfigure[\;$\gamma = 0.3$, $v_0=0.1$]{\includegraphics[width=7.0cm,clip=]{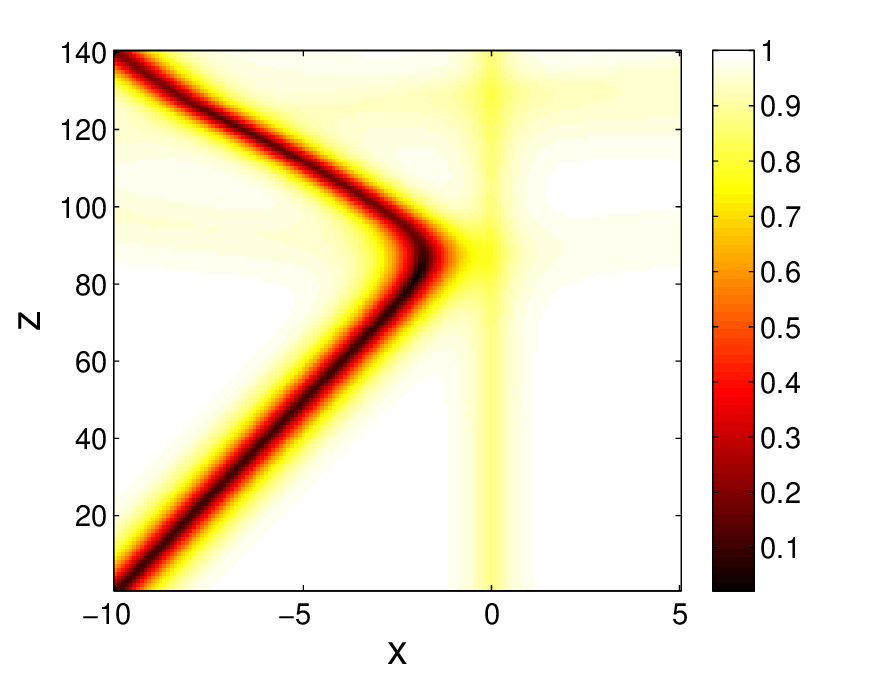}}   \hspace{1cm}
\subfigure[\;$\gamma = 0.3$, $v_0=0.8$]{\includegraphics[width=7.0cm,clip=]{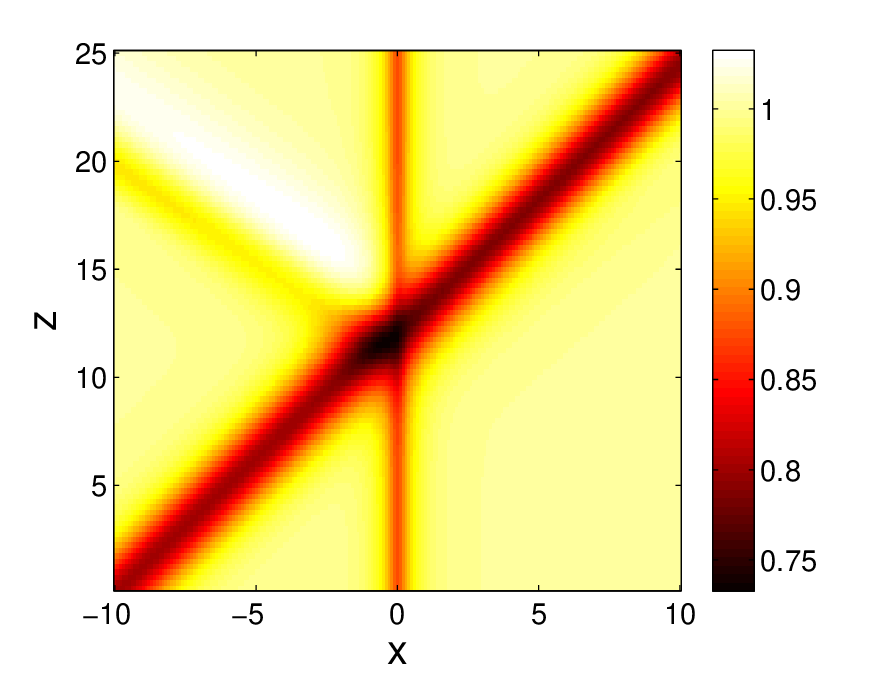}} 
\subfigure[\;$\gamma =-0.3$, $v_0=0.1$]{\includegraphics[width=7.0cm,clip=]{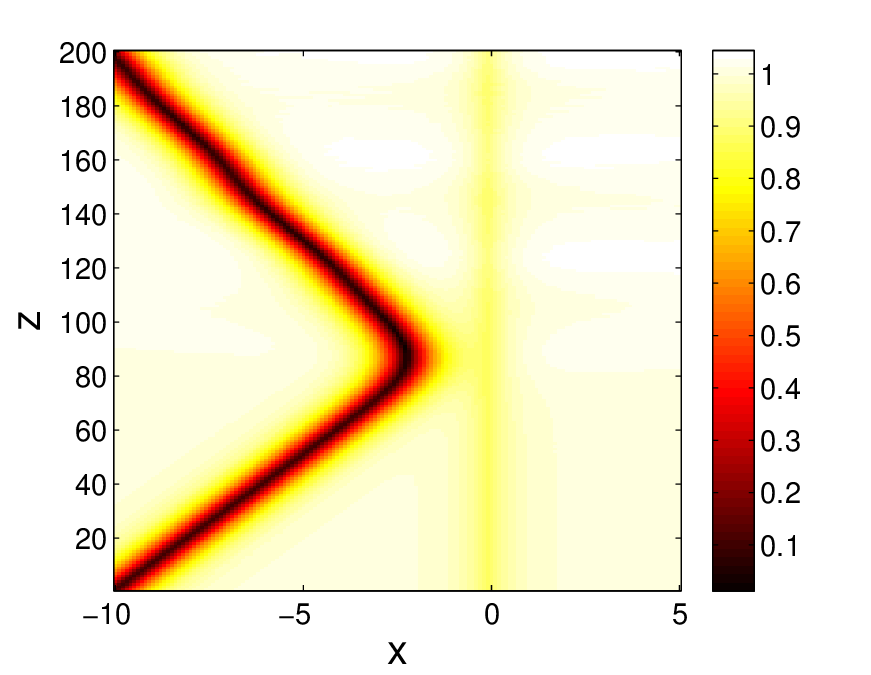}} \hspace{1cm}
\subfigure[\;$\gamma =-0.3$, $v_0=0.8$]{\includegraphics[width=7.0cm,clip=]{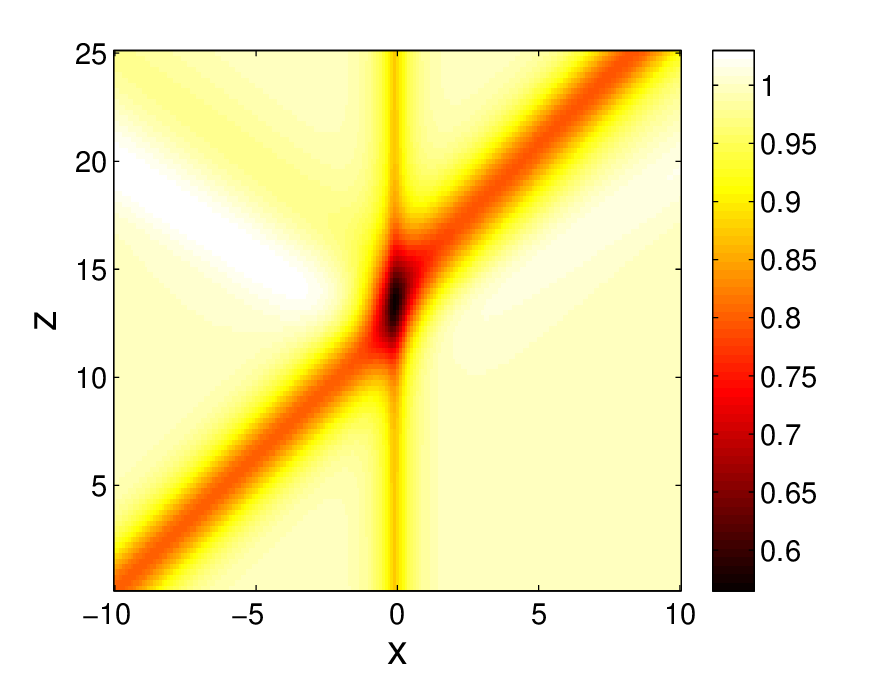}}
\caption{(Color online) The interaction of dark solitons with the $\mathcal{%
PT}$-symmetric dipole at values of the parameters indicated in the panels,
and $\protect\epsilon =0$. Shown is the top view of the intensity
distribution, $|\protect\psi (x,z)|$.
(a) $\gamma =  0.3$, $v_0 = 0.1$; (b) $\gamma =  0.3$, $v_0 = 0.8$;
(c) $\gamma = -0.3$, $v_0 = 0.1$; (d) $\gamma = -0.3$, $v_0 = 0.8$.}
\label{ds1}
\end{figure*}

\subsection{Dynamics of bright solitons}

In the model with the self-focusing nonlinearity, $g=+1$ in Eq. (\ref{NLS}),
we simulated collisions of the incident bright soliton with the $\mathcal{PT}
$ dipole, setting $\epsilon =0$ in Eq.\ (\ref{delta}). The initial
conditions are taken as per expression (\ref{sol}), i.e.,
\begin{equation}
\psi \left( x,0\right) =\eta ~\mathrm{sech}\left[ \eta (x-x_0\right] \exp
\left( iv_0\left(x-x_0\right)\right),
\end{equation}%
centered at $x_{0}=-10$, with initial velocity $v_{0}>0$, and $\eta =1$
(once $\epsilon =0$ was set, $\eta =1$ may be always fixed by rescaling).

Shown in Fig.\ \ref{bs1} are two pairs of examples of the interaction of the
soliton with the dipole. In panels (a,b), the case of $\gamma >0$ is
considered, which, according to Eqs. (\ref{NLS}), (\ref{delta}) and (\ref%
{deltas}), implies that the incident soliton impinges on the dipole from the
side where the amplifying (rather than attenuating) element is located. In
panel (a) of Fig.\ \ref{bs1}, the soliton gets trapped by the defect and subsequently
blows up, which happens when the initial velocity is sufficiently small. On
the other hand, when the velocity is sufficiently large, the incoming
soliton, quite naturally, passes the defect, as seen in panel (b). These two
examples are typical for such outcomes of the collision.

In panels (c,d) of Fig.\ \ref{bs1}, we display the evolution of the soliton
for $\gamma =-0.5$, when the the incident soliton approaches the dipole from
the side of the attenuating element. On the contrary to panel (a), where trapping followed by the
blowup was observed, in the present case the incident soliton is reflected
if its velocity is small enough. Naturally, the reflected soliton has a
smaller amplitude than the incident one, due to the action of the
attenuation. On the other hand, it is shown in panel (d) the the soliton
passes the defect if the velocity is large enough, similar to what was
observed for $\gamma >0$ in panel (b). In all the panels, the insets show
the reflected and transmitted powers, $P_{R}/P_{I}$ and $P_{T}/P_{I}$,
defined according to Eq. (\ref{powers}), as above. Their asymptotic values
at $z\rightarrow \infty $ are compared with the exact results, $(R_{1}^{2} + R_{2}^{2})$ 
and $(T_{1}^{2} + T_{2}^{2})$, for the linear plane waves, as
given by Eq. (\ref{RT}).

Obviously, an important characteristic of the interaction of the soliton
with the $\mathcal{PT}$ dipole, which also includes the attractive or
repulsive local potential, as per Eq. (\ref{delta}), is the minimum
(threshold) velocity necessary for the soliton to pass thes defect 
(possibly, plassing in the form of a pulse which is not exactly a soliton). 
We aim to identify the threshold velocity produced by the direct simulations of
Eq.\ (\ref{NLS}), and compare it to predictions of the semi-analytical
approximation based on quasi-particle equations\ (\ref{d/dz}), (\ref{d/dz2}), 
and (\ref{dv/dz}). Because not the entire power is reflected or
transmitted as a result of the collision, we define the soliton as being
transmitted past the defect when, at least, half of its total power is
transmitted, i.e., in terms of the insets of Fig.\ \ref{bs1}, the
transmission threshold corresponds to the point where the solid blue and red
lines cross.

The threshold velocities produced by the direct simulations are displayed in
Fig.\ \ref{bs3}, as functions of the $\mathcal{PT}$-dipole's strength, $\gamma$, 
by crosses. Predictions produced by a numerical solution of Eqs.~(\ref{d/dz}) 
and (\ref{d/dz2}) are presented too in this figure, by means of
dashed lines. In addition to the approximation based on Eqs.\ (\ref{d/dz})
and (\ref{d/dz2}), for the case of small $\epsilon $, such as in panels (b)
and (d), we also plot, by dashed-dotted lines, the prediction generated by
the numerical solution of Eqs.\ (\ref{d/dz}) and (\ref{dv/dz}), which was
derived for $\epsilon =0$.

It is seen from Fig. \ref{bs3}(a) that, as mentioned in Section \ref{sec2}%
.B, the quasi-particle adiabatic approximation, based on Eqs. (\ref{d/dz})
and (\ref{d/dz2}), is in good agreement with the direct simulations for $%
0<\gamma <\epsilon $. On the other hand, panels (b) and (d) demonstrate that
the \textit{post-adiabatic} approximation, represented by Eqs. (\ref{d/dz})
and (\ref{dv/dz}), which takes into regard the generation of the intrinsic
chirp in the soliton [see Eq. (\ref{b})], is relevant for $|\epsilon |\ll
\gamma \lesssim 0.2$ (which implies that $\gamma $ is positive). The same
approximation provides for a qualitative prediction of the threshold
velocity for $\gamma <0$ and $|\epsilon |\ll |\gamma |$ too, even though in
that case the prediction is quantitatively inaccurate. The reason is that,
as can be seen from numerical data (not shown here in detail), the
deformation of the soliton around $x=0$ is not small in the latter case,
which cannot be taken into account by the perturbative treatment. In this
sense, a better agreement with the perturbation theory may be expected for a
smoother shape of the $\mathcal{PT}$ dipole [see Eq. (\ref{deltas})], but in
that case the analytical results take a more cumbersome form, due to the
complexity of the respective integrals in Eqs.\ (\ref{gamma}), (\ref{dP/dz}%
), and (\ref{dv/dz}). Generally, the fact that the discrepancy between the
numerical and analytical results in Fig.\ \ref{bs3} is smaller for $\gamma >0
$ is explained by the fact that the larger amplitude of the pumped, rather
than attenuated, soliton in this case (see above) makes the local
perturbation weaker in comparison with other terms in Eq.\ (\ref{NLS}).

\subsection{Dark solitons}

To consider the interaction of dark solitons with the $\mathcal{PT}$ dipole,
we fix the CW-background amplitude in Eq.\ (\ref{dark}) as $\mu =1$. In the
absence of dark solitons, the CW background, $\psi _{\mathrm{CW}}$, is
deformed by the potential \cite{konotop04,sakaguchi05,li11,achi12}. 
As shown above, in the limit of $\epsilon =0$ and ideal $\delta^{\prime}$ 
function in Eq.~(\ref{dark}), the deformation amounts to the phase jump (\ref{shift}) at $x=0$.

In Fig.\ \ref{bg}(a), we plot the shape of the background obtained in the
numerical form, with the $\delta ^{\prime }$ function in Eq.\ (\ref{dark})
replaced by regularization\ (\ref{deltas})], for $\gamma =0.3$ and $\epsilon
=0$. Similarly to the previous works, we find that this ground state,
produced by the stationary solution of Eq.\ (\ref{dark}), exists at $\gamma
<0.49$ (at $\gamma $ exceeding this critical value, the system starts
spontaneous generation of dark solitons \cite{li11,achi12}).
The difference of the background from the above-mentioned analytical
solution, which amounts to the phase jump (\ref{jump}) embedded into the
constant background, is explained by the difference of approximation (\ref%
{deltas}) from the ideal $\delta ^{\prime }$ function. Additionally, we also
plot in the same figure in panel (b) and (c) the profile of the plane waves
in the presence of nonzero $\epsilon$.


Due to the presence of the non-uniform CW background ($\psi _{\mathrm{CW}}$), 
we simulated the dynamics of a dark soliton in the framework of Eqs.~(\ref{dark}) and~\eqref{deltas} with initial conditions
\begin{widetext}
\begin{equation}
\psi(x,0)=\psi _{\mathrm{CW}}\left[ \sqrt{1-v_{0}^{2}}\tanh \left( \sqrt{%
1-v_{0}^{2}}(x-x_{0})\right) +iv_{0}\right] ,  \label{initial-dark}
\end{equation}%
\end{widetext}
where $v_{0}$ and $x_{0}$ determine the initial velocity and position of the
dark soliton.

In Fig.\ \ref{ds1}, we plot simulated pictures of the interaction of the
dark soliton with the $\mathcal{PT}$ dipole for parameter values indicated
in the caption to the figure (cf. pictures for the interaction of dark solitons with conservative local defects in Ref.~\cite{kono97}). 
Similar to the case of bright solitons considered above, the dark soliton is either transmitted or reflected.
In panel (b) of Fig. \ref{ds1}, an extra weaker reflected dark soliton
emerges too, as a result of the interaction, in addition to the main passing
soliton. Another particular result is seen in panel (d), where the reflected
feature, observed at $x<0$, is not a soliton but a broad shallow perturbation, propagating with
the speed determined by the background amplitude (the generation of such perturbations
by dark solitons was considered in Ref.~\onlinecite{ablo11}).

The analytical approximation for the dark-soliton dynamics, based on
Eqs.~(\ref{kappa}) and~(\ref{zeta}) for variables $\kappa (Z)$ and $\zeta
(Z)$, was derived in the framework of the adiabatic approach, which does not
take into regard the generation of the additional dark soliton and shallow perturbation, 
hence this approximation cannot describe the observed phenomenology accurately. 
Nevertheless, predictions of the analysis may
qualitatively explain some features of the dynamics revealed by numerical
simulations. For the sake of the comparison, obtaining coordinate $\zeta$
from results of the simulations of Eq.~(\ref{dark}) is straightforward, while
amplitude $\kappa $ can be identified as 
$\kappa(z) = \mathrm{sign}(x_{0}) \sqrt{\left(1 - |\psi(x = \zeta,z)|^{2} \right)/2}$. 
Note also that the analytical approximation was derived under the assumption of $|v_{0}| \sim 1$. 
In that regard, the approximation may only be compared with the dynamics displayed in
panels (b) and (d). In particular, in the case shown in Fig.\ \ref{ds1}(b),
the approximation correctly predicts that the incident dark soliton would
pass through the $\mathcal{PT}$ dipole, although there is a discrepancy in
approximating the phase shift of the soliton after the interaction--most
plausibly, caused by the fact that the adiabatic approximation does not take
into account the generation of the additional reflected soliton, in this
case. Nevertheless, the approximation correctly predicts that the soliton
accelerates in the vicinity of the dipole.


\section{Conclusion}

\label{sec4}

We have studied the dynamics of bright and dark solitons in the model based
on the focusing and defocusing NLS equations with an embedded defect in the
form of the $\mathcal{PT}$-symmetric dipole, combined with a local repulsive
or attractive potential. The scattering problem for plane waves and broad
incident packets was considered too in the framework of the linear version
of the model.
An essential difference from previously studied interactions of solitons and linear waves with defects in conservative systems is that,
in spite of the gain-loss balance in the $\mathcal{PT}$-symmetric dipole, the collisions change the norm, i.e.,
effective mass, of the solitons and wave packets, making the interaction dynamics more complex.
In particular, the basic dynamical system approximating the collision for the bright solitons, 
based on Eqs.~\eqref{d/dz} and~\eqref{d/dz2}, is of the third order, instead of the well-known second-order approximation in conservative systems.
The numerical study for the focusing nonlinearity has produced
threshold values of the velocity of the incident bright soliton above which
it passes the local defect. For the defocusing nonlinearity, the interaction
of dark solitons with the defect is studied in the numerical form too.
Parallel to the simulations, we have developed analytical approximations for
both cases. For the bright solitons, the adiabatic quasi-particle
approximation yields accurate results in the case when the repulsive
potential is stronger than the gain-and-loss component of the defect. For
the negligibly weak local potential, the analytical consideration goes
beyond the limit of the adiabatic approximation, taking into regard the
intrinsic chirp of the soliton. The respective semi-analytical results
predict the threshold velocity in a reasonably accurate form too. For the
dark solitons, the approximation qualitatively explains the transmission,
acceleration and deceleration of the incident soliton. In addition, the
exact solution for the dark soliton pinned by the $\mathcal{PT}$-symmetric
defect was found too. \newline

\section*{ACKNOWLEDGEMENTS}

W.H. and H.S. thank the School of Mathematical Sciences, University of
Nottingham, for a Summer Research Bursary (June--September 2013).
N.K. acknowledges the Seed Grant K$\Phi$-13/28 from the ``Fund of Social Development''
Corporate Fund and Central Research Office (CRO) of Nazarbayev University Research
and Innovation System (NURIS), Kazakhstan.

\end{document}